\documentstyle[12pt,aasms4]{article}
 
\newcommand{\etal}{{\it et al.\/}}
\newcommand{\ie}{{\it i.e.}}

\lefthead{Sigut, Landstreet \& Shorlin}
\righthead{Emission from 3 Cen A}
 
\begin{document}

\renewcommand{\thefootnote}{\fnsymbol{footnote}}

\title{Emission Lines in the Spectrum of the $\bf ^3$He Star 3 Cen A
\footnote{Based on observations obtained with the
Canada-France-Hawaii telescope, operated by the National
Research Council of Canada, the Centre National de Scientifique
of France, and the University of Hawaii.}}

\author{T.\ A.\ A.\ Sigut, J.\ D.\ Landstreet and 
S.\ L.\ S.\ Shorlin}
\affil{Department of Physics and Astronomy,
The University of Western Ontario}
\authoraddr{London, Ontario, Canada N6A 3K7}

\slugcomment{Accepted for publication by the 
Astrophysical Journal Letters}

\begin{abstract}

Emission in the $\rm 4d\rightarrow 4f$ transitions of \ion{Mn}{2} 
(multiplet 13,
$\lambda\lambda\,6122-6132\,$\AA), in the $\rm 4f\rightarrow 6g$ transitions of \ion{P}{2},
and in $\lambda\,6149.5$ of \ion{Hg}{2} has been detected in the
spectrum of the helium weak star 3~Centauri~A (B5 III-IVp).
Weaker emission from the same \ion{Mn}{2} multiplet is also seen in the hot, mild
HgMn star 46~Aquila (B9 III).
It is suggested that the emission is of photospheric origin
and may be evidence for the stratification of manganese,
phosphorus and mercury in the photosphere of 3~Cen~A, and of 
manganese in 46~Aql.

\end{abstract}


\section{Introduction}

3~Centauri~A (HD~120709)
is a B5 III-IVp He-weak star of the Ga-P subclass. Its spectrum is
characterized by a highly non-solar chemical composition, with
large overabundances (1-2~dex) of P, Sc, Mn, and Cu, extreme
overabundances (3-4~dex) of
Ga, Kr, and Hg, and underabundances (0.5-1 dex) of He, B, C, N, Mg,
and Al (Castelli, Parthasarathy \& Hack 1997). In addition, most of
the helium in its photosphere is in the form of $^3$He, with a
$\rm ^3He/^4He$ ratio of $\approx 2.5$ as compared to the solar ratio of 
$1.4\cdot10^{-4}$ (Sargent \& Jugaku 1961, Hartoog \& Cowley 1979).
These abundance peculiarities are thought to have their ultimate explanation in terms of
elemental diffusion operating in the photosphere. The
mechanism of light-induced-drift seems capable of explaining the
very large $^3$He to $^4$He ratio (Michaud \& Proffitt 1992, LeBlanc \& Michaud 1993).
No longitudinal magnetic field has been detected in 3~Cen~A to an 
upper limit of $\sim 100\,$G (Borra, Landstreet
\& Thompson 1983). Chemical abundances derived from
optical spectra have been recently reported by Pintado, Adelman, \& Gulliver (1998)
and UV spectral synthesis has been performed by Castelli \etal\ (1997). 

In this {\it Letter\/},
we add to the peculiarities of 3~Cen~A by reporting the presence of
several emission lines of \ion{Mn}{2} (multiplet~13) and \ion{P}{2}, 
and one of \ion{Hg}{2} ($\lambda\,6149.2$), in its red spectrum.
We also report emission in the same \ion{Mn}{2} multiplet in
the spectrum of the prototype hot, mild\footnote{Mild in this context
refers to the relatively moderate manganese photospheric 
abundance enhancement.} HgMn star 46~Aquila (HD~186122).

\section{Observations}

Spectra are available for 3~Cen~A in two wavelength regions,
$\lambda\lambda\,6105-6190$ and $\lambda\lambda\,6350-6425$, which were taken with the
coud\'{e} f/4 (Gecko) spectrograph at the Canada-France-Hawaii telescope in May 1999.
These observations were obtained as part of a program to study
neon abundances in late and mid B stars in an attempt to detect very
weak stellar winds, as 
suggested by Landstreet, Dolez \& Vauclair (1998). The CCD detector EEV2 was used,
giving 200\,x\,4500 pixels, each with a size of 13~$\mu m$.
The spectra were bias subtracted, flat-fielded, and continuum normalized using the IRAF
reduction package (Tody 1993). The spectra were wavelength calibrated
from a ThAr comparison
arc spectrum taken after each grating setting; the resulting dispersion relation limits the
relative wavelengths to an accuracy of $\pm 0.003\,$\AA.
The Julian dates of each observation, as well as
other details, are summarized in Table~\ref{tab:obs}.
As the Gecko spectrograph offers a resolution of $\sim 1.2\cdot 10^5$, it is
a superb instrument with which to study ultra-sharp-lined stars such as 3~Cen~A.

In the following discussion, the spectra of 46~Aql (B9~III HgMn, HD~186122) and
$\kappa$~Cnc (B8 IIIp HgMn, HD~78316),
obtained during the same observing run, will be used for comparison with
the spectra of 3~Cen~A; their
observational details are also summarized in Table~\ref{tab:obs}.

\section{Results}

Figure~\ref{fig:main} shows the spectrum of 3~Cen~A in both spectral regions.
Prominent absorption lines dues to \ion{O}{1}, \ion{Si}{2}, \ion{Fe}{2}, \ion{Ne}{1},
and \ion{P}{2} are indicated. The final wavelength correction for the radial velocity
of 3~Cen~A was made using the laboratory wavelengths of these strong absorption lines.
After applying a mean shift, the wavelengths of these reference lines
were reproduced to within $\pm 0.01\,$\AA.
As no variability was detected between observations on successive nights,
all eight 300 second exposures for each wavelength region of
3~Cen~A listed in Table~\ref{tab:obs} were combined.

\subsection{\protect\ion{Mn}{2} Emission}

Figure~\ref{fig:mn2} shows the spectral region of
$\lambda\lambda\,6120-6135$ which contains
\ion{Mn}{2} multiplet~13, $\rm 3d^5(^6S)4d\,^5D \rightarrow
3d^5(^6S)4f\,^5F^o$.
These are transitions between highly excited states, with the
lower levels lying $\sim 10.2~$eV above ground. In the Figure, the multiplet
structure is shown on top, with the individual line lengths proportional to
the $gf$ values of the transitions. The required atomic data for this multiplet were
obtained from the Kurucz \& Bell (1995) database. As can be seen, the entire multiplet structure
is accounted for and the identification is not in doubt. For reference, the predicted
LTE spectrum for 3~Cen~A using the abundances of Pintado \etal\ (1998), with 
a manganese enhancement of 1.5~dex,
and a Kurucz model
atmosphere of parameters $T_{\rm eff}=17\,500\;$K and $\log(g)=3.8$, is shown in
the figure.  The spectrum
of the HgMn star $\kappa$~Cnc in the same wavelength range is also shown, which
presents this \ion{Mn}{2} multiplet in strong absorption. Note that $\kappa$~Cnc is
significantly cooler than 3~Cen~A with atmospheric parameters of
$T_{\rm eff}=13\,470\;$K and $\log(g)=3.76$ as derived by Smith (1992).

3~Cen~A is not the only star in our sample to show 
emission in multiplet 13 of \ion{Mn}{2}.
Weak emission in this multiplet is also definitely detected in
the spectrum of 46~Aql (B9 III) which is the prototype of the
hot, mild HgMn stars (Cowley 1980, Smith 1993). Its spectrum is also shown in
Figure~\ref{fig:mn2}. Smith (1992) derives atmospheric parameters for 46~Aql of
$T_{\rm eff}=13\,000\;$K and $\log(g)=3.65$, making it significantly cooler than
3~Cen~A and comparable to $\kappa$~Cnc.

For 3~Cen~A, the observed wavelengths of the \ion{Mn}{2} emission lines
have a mean shift relative to the photospheric absorption lines of $\sim 0.015\;$\AA,
at the limit of the wavelength calibration. Hence, there is no evidence that
the emission is Doppler shifted relative to the photospheric absorption lines.
The lines in the 3~Cen~A spectra are resolved and symmetric, and the FWHM of the emission
features are not noticeably different from the absorption lines.

The relative intensities of the emission lines in the \ion{Mn}{2} multiplet
in both 3~Cen~A and 46~Aql are not
exactly proportional to their $gf$ values, suggesting optically thick, not
optically thin, formation. Note that the laboratory spectrum of this multiplet
presented by Johansson \etal\ (1995) 
gives relative intensities in excellent agreement
with those predicted by the Kurucz \& Bell (1995) $gf$ values, 
suggesting that the $gf$ values have high relative accuracy, as is frequently the case
for Kurucz's calculations within a single multiplet.

There are no traces of absorption components to the emission lines,
\ie\ none of the emission lines sit on an absorption trough. Hence either
the emission has completely filled in the photospheric absorption components
expected to be present in LTE (see later discussion) 
due to the high Mn abundance of 3~Cen~A derived from
other wavelengths (see Figure~\ref{fig:mn2}), 
or the emission {\em is\/} the photospheric spectrum.

\subsection{\protect\ion{Hg}{2} and \protect\ion{P}{2} Emission}

Figure~\ref{fig:hg2} shows the spectral region
region $\lambda\,6144-6160$ of 3~Cen~A which contains two prominent
\ion{Fe}{2} transitions of multiplet 74, $\lambda\,6147.7$ and $\lambda\,6149.2$.
The emission feature close to the \ion{Fe}{2} line
at $\lambda\,6149.2$ is definitely identified with \ion{Hg}{2} $\lambda\,6149.5$. 
Identification
of this feature, and its impact on using the relative strengths of these two
\ion{Fe}{2} lines to detect
magnetic fields, is discussed by Takada-Hida \& Jugaku (1992) and 
Hubrig, Castelli \& Wahlgren (1999).
In Figure~\ref{fig:hg2},
the spectrum of $\kappa$~Cnc is shown for reference, and the blend of
\ion{Hg}{2} in the wing of the \ion{Fe}{2} line is clearly seen.

We note that the apparent increase in strength of \ion{Fe}{2} $\lambda\,6147.7$ over
that of $\lambda\,6149.2$ is probably due to a blend 
in the red wing of $\lambda\,6147.7$ with
another (high-excitation) transition of \ion{Fe}{2} and thus there is no
evidence for a magnetic field in 3~Cen~A using the technique of Mathys \& Lanz (1990). 
However, the ultra-sharp lines of 3~Cen~A can be used to constrain
any non-thermal broadening due to rotation, microturbulence, or a magnetic field. 
The largest possible
value for each (assuming the other two are negligible) consistent with the line widths
have been determined by modeling the \ion{Fe}{2} and \ion{Ne}{1} line profiles.
The resulting limits
are $v\,\sin\,i\le 2\;\rm km\,s^{-1}$, $\zeta_t \le 1.5\;\rm km\,s^{-1}$, and
$\left< B_s \right> \le 1500\;$G (where $\left< B_s \right>$ is the mean field modulus
averaged over the visible hemisphere).

Many of the remaining weak emission features in the observed spectrum of 3~Cen~A
can be identified with high-excitation \ion{P}{2} transitions, having as their
lower levels symmetries of the $\rm 3s^2\,3p\,(^2P^o)\,4f$ configuration near
16.2~eV, and their upper levels, symmetries of the $\rm 3s^2\,3p\,(^2P^o)\,6g$
configuration.
Several \ion{P}{2} transitions are also identified in the second
spectral window of $\lambda\lambda\,6350-6425$. These identifications have been made with
the atomic line list of van Hoof\footnote{http://www.pa.uky.edu/$\sim$peter/atomic/} which
is based on the energy level data available in the NIST Atomic Spectroscopic Database.

The measured wavelengths of all of the emission features in both spectral
windows are given in Table~\ref{tab:emiss}, along with their identifications
and $\log(gf)$ values, where available. A few weak emission features
still remain unidentified. In this latter group, we have noted a few
wavelength co-incidences with permitted transitions of
\ion{Si}{1} from the $\rm 3s^23p\,4p$ configuration. However, as these transitions
are all to highly excited states, $n>10$, and as 
3~Cen~A shows absorption lines of only \ion{Si}{2} and
\ion{Si}{3} in its optical and UV spectrum, consistent with its $T_{\rm eff}$, 
we regard these identifications as extremely unlikely. Given 3~Cen~A's P-Ga classification,
emission lines due to \ion{Ga}{2} may be a interesting possibility, but the energy
level identifications of \ion{Ga}{2} are incomplete and no line lists of \ion{Ga}{2}
transitions in the red ($\lambda>6000\;$\AA) are currently available.

\section{Formation Mechanisms}

The emission lines described in the previous section appear to be photospheric,
not circumstellar, in origin. The lines are resolved and symmetric, and their FWHM are similar
to the absorption lines present in each spectrum. The radial velocity derived from the
absorption and emission lines agrees to within the uncertainty in the wavelength calibration.
%
%

Weak emission lines of photospheric origin are not unknown. Vega exhibits
weak emission lines of \ion{Fe}{2} in the wings of Lyman-$\alpha$ 
(van Noort et al.\ 1998) and weak emission lines from iron and rare earth elements 
appear in the wings of the \ion{Ca}{2} H \& K lines in the solar spectrum
(Cram, Rutten \& Lites 1980). In both instances, it is strong wing opacity
in Lyman-$\alpha$ or \ion{Ca}{2} which shifts the depth of formation of the weaker lines
to small optical depths where interlocked non-LTE effects produce
source functions which rise with height (decreasing optical depth).
Interlocked non-LTE effects also produce the well known solar \ion{Mg}{1} 
12~$\mu m$ lines (Carlsson, Rutten \& Shchukina 1992). 
None of these mechanisms require 
a temperature rise in a chromosphere to produce the emission.

\ion{Mn}{2} multiplet 13 is clearly seen in emission in the spectrum of
$\eta$ Carin\ae\ (Johansson et al.\ 1995); however, this spectrum is
of circumstellar origin. Johansson \etal\ suggest that this \ion{Mn}{2} emission
is formed by fluorescent excitation via a wavelength coincidence of \ion{Mn}{2}
multiplet 15 and \ion{Si}{2} multiplet UV 5. However, we are doubtful
that a fluorescent origin for the \ion{Mn}{2} lines is possible in the present case.
For fluorescent excitation to occur, the source function of the pumping
transition must be different from the Planck function {\em in the
line forming layers}, which, in the current case, would be deep in the
photosphere where the collisional rates are large (although perhaps chemical
stratification could play a role - see later discussion).

A non-LTE origin for these emission lines is suggested as only
certain transitions for each ion are in emission. Both absorption and
emission lines of \ion{P}{2} are present in our spectra, and \ion{Mn}{2} and \ion{P}{2}
absorption lines appear in optical and UV spectra forming the basis for the
the published abundance studies. Interlocked non-LTE
effects are capable of producing line source functions that rise with
height (or decreasing optical depth) without the requirement of a chromospheric
temperature rise. As $h\nu<kT$ for transitions at the wavelengths
considered, photon losses in the lines themselves are insufficient to
significantly influence the departure coefficients of their upper and lower levels.
The departure coefficients can be controlled by losses in other lines,
or bound-free rates, which can produce rising source functions in the
same manner as for the solar \ion{Mg}{1} 12~$\mu$m lines (Carlsson \etal\ 1992). 
As this mechanism is transition specific, it does not follow that
the entire manganese, phosphorus, or mercury spectrum need be in emission.

There is an additional interesting point to consider:
the large overabundances of manganese, phosphorus, and mercury
are thought to result from diffusive processes which will act
to concentrate these elements in thin layers at small optical depths
(Michaud, Reeves \& Charland 1974, Alecian \& Michaud 1981).
If this is the case, then for example, the entire manganese or mercury spectrum would
be produced at small optical depths where the collisional rates are very small
and non-LTE effects can dominate. The chemical stratification would then play
a role similar to the overlapping wing opacity in the previous cases cited.
While this mechanism is plausible, detailed non-LTE calculations are
required to confirm this hypothesis, and in particular, to determine if
chemical stratification of the abundance distribution plays a critical
role. Such calculations are planned for the near future. Additional observations are
required to capture more of the \ion{Mn}{2} spectrum. If our hypothesis
is indeed correct, these emission features may provide rather direct
evidence of chemical stratification in the photosphere of 3~Cen~A
providing addition constraints on the diffusive model for 
chemically peculiar stars.

\section{Conclusions}

We have discovered weak emission due to \ion{Mn}{2}, \ion{Hg}{2} and \ion{P}{2}
in the spectrum of 3~Cen~A
in the wavelength regions $\lambda\lambda\,6105-6190$ and $\lambda\lambda\,6350-6425$.
We have also discovered \ion{Mn}{2} emission in the spectrum of the hot, mild
HgMn star 46~Aql.
We suggest that these emission lines are of photospheric origin and are produced
by interlocked non-LTE effects. As the elements observed to produce the emission
are very probably made overabundant by chemical diffusion, it is speculated that the
concentration of absorbers in thin layers at small optical depth may play an
important role in the formation of the emission lines.
Additional observations of more He-weak and HgMn stars, as
well as wider wavelength coverage for 3~Cen~A, are required.
A more complete survey of these spectral regions, for both normal and chemically
peculiar stars, would be particularly valuable.

\acknowledgements
TAAS wishes to thank J.\ M.\ Marlborough and
J.\ D.\ Landstreet for support through their NSERC grants. We particularly
thank Elizabeth Griffin for suggesting
the \ion{P}{2} and possible \ion{Si}{1}
identifications. We acknowledge the
assistance of Simon Strasser with the IRAF reductions.

\newpage

\figcaption[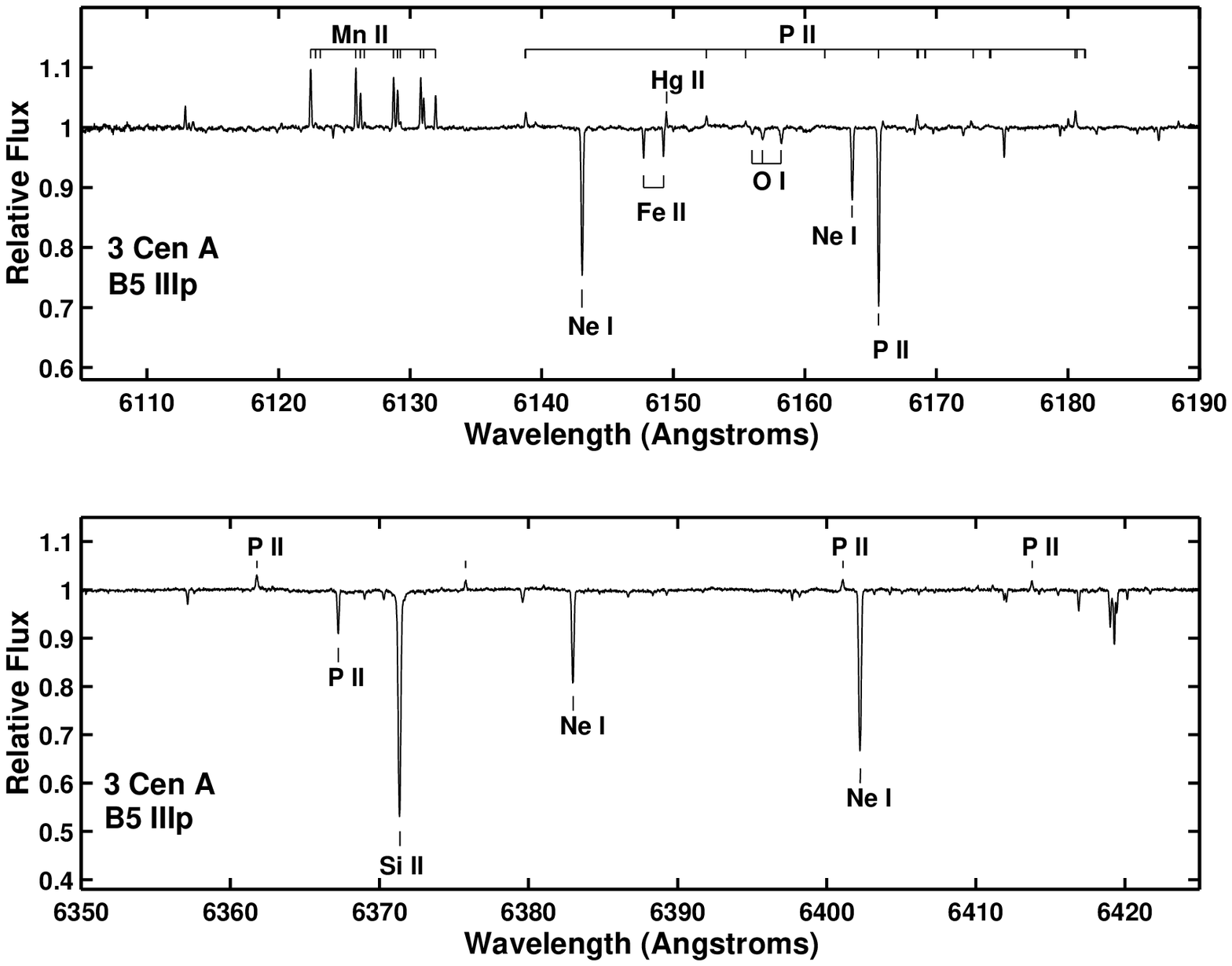]{\small The observed spectra of 3~Cen~A with
all available observations combined. The emission lines 
and their identifications are noted, along with some prominent absorption lines.
\label{fig:main}}

\figcaption[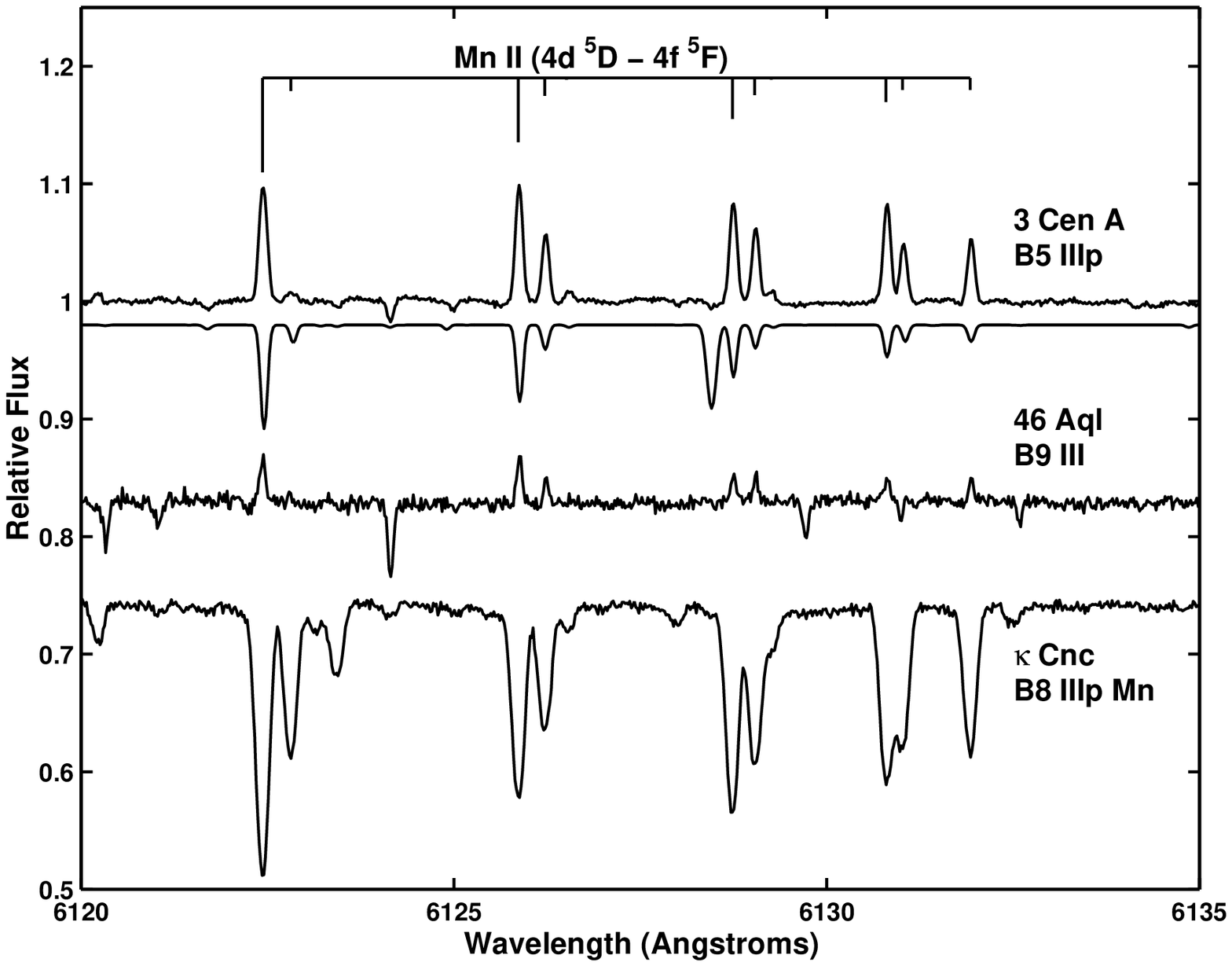]{\small A close-up of the wavelength 
region around \protect\ion{Mn}{2} multiplet 13.
In the top multiplet identification,
the length of each line is proportion to the
$gf$ value of that transition. The spectra plotted are, from top to bottom,
the observed spectrum of 3~Cen~A, the theoretically predicted spectrum 
for 3~Cen~A using a Kurucz model
atmosphere (see text for details), the spectrum of 46~Aql, and the spectrum of
$\kappa$~Cnc. \label{fig:mn2}}

\figcaption[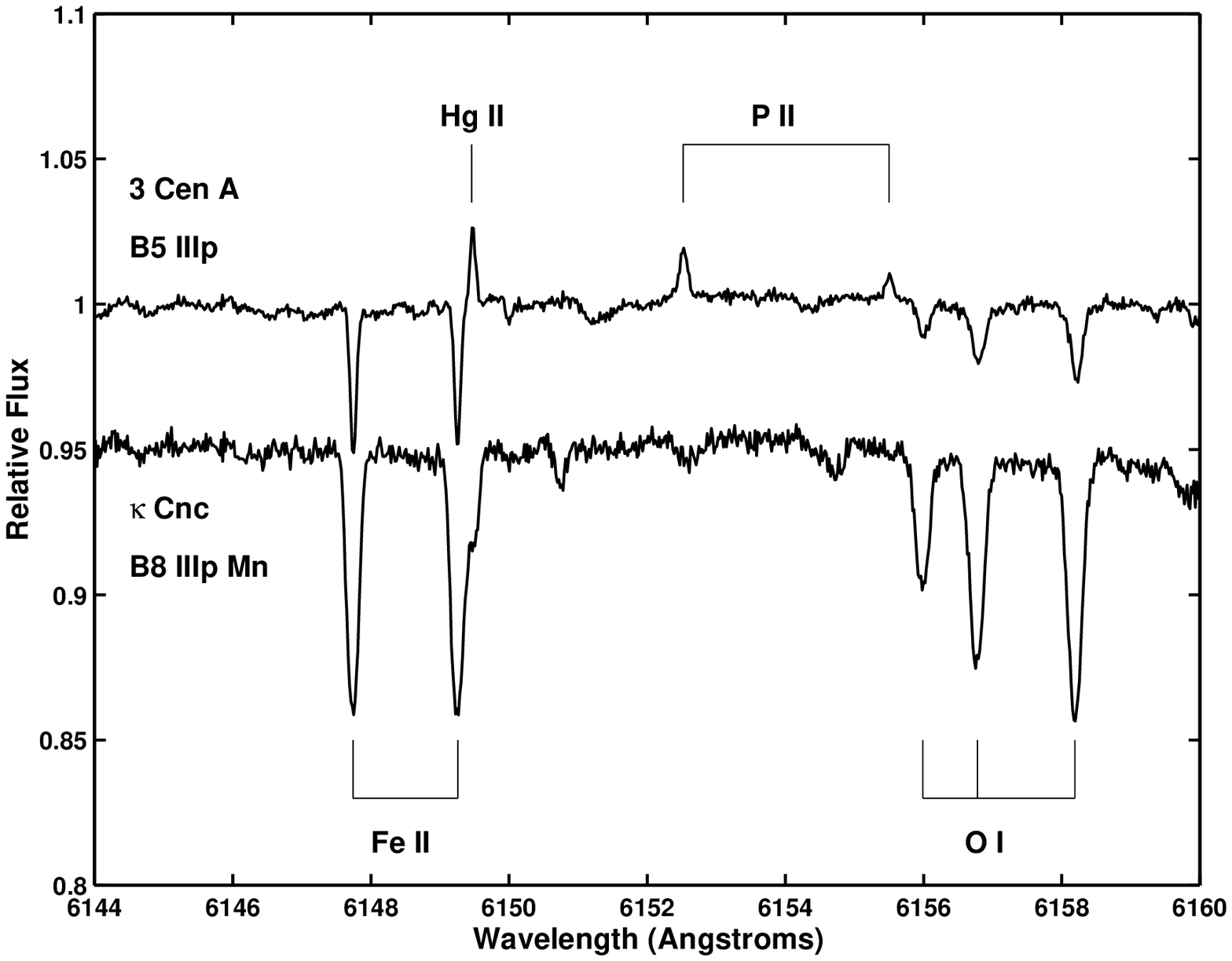]{The \protect\ion{Hg}{2} emission line at $\lambda\,6149.5\,$.
The top spectrum is that of 3~Cen~A while the spectrum of
$\kappa$ Cnc is offset at the bottom. \label{fig:hg2}}

\begin{deluxetable}{llcccc}
\small
\tablewidth{0pt}
\tablecaption{Stellar Observations.
\label{tab:obs}}
\tablehead{
\colhead{Star} & \colhead{Spectral Class} & \colhead{JD} &
\colhead{Wavelength} & \colhead{Exposure} & \colhead{$\rm S/N$} \\
  &   &   &  (\AA)  & (sec) & }
\startdata
3~Cen~A & B5 III-IVp & 2451325.7875  & 6105 - 6190 & $\rm 4x\,300$ & 550 \nl
        &            & 2451326.7998  & 6105 - 6190 & $\rm 4x\,300$ & 600 \nl
3~Cen~A & B5 III-IVp & 2451324.7752  & 6350 - 6425 & $\rm 4x\,300$ & 500 \nl
        &            & 2451326.8247  & 6350 - 6425 & $\rm 4x\,300$ & 450 \nl
        &            &          &             &       &     \nl
$\kappa$ Cnc & B8 IIIp HgMn & 2451325.7230 & 6105 - 6190 & $\rm 4x\,300$ & 400 \nl
             &              & 2451325.7230 & 6350 - 6425 & $\rm 4x\,300$ & 375 \nl
46 Aql & B9 III HgMn        & 2451326.0950 & 6105 - 6190 & $\rm 4x\,300$ & 300 \nl
             &              & 2451327.0550 & 6350 - 6425 & $\rm 4x\,300$ & 350 \nl
\enddata
\end{deluxetable}

\begin{deluxetable}{rrlrr}
\small
\tablewidth{0pt}
\tablecaption{Emission Features in the spectrum of 3~Cen~A. 
\label{tab:emiss}}
\tablehead{
\colhead{$\lambda^{\rm obs}$} & \colhead{EW} &
\colhead{Element} & \colhead{$\lambda^{\rm exp}$} &
\colhead{$\log(gf)$} \\
\colhead{(\AA)}  & \colhead{(m\AA)} &      &  \colhead{(\AA)} & }
\startdata
6112.90 &  4 &  \nodata       & 6112.88\phn  &\nodata \nl
6113.18 &  7 &  \nodata       & 6113.24\phn  &\nodata \nl
6113.50 & 12 &  \nodata       & 6113.47\phn  &\nodata \nl
6122.43 & 13 &  \ion{Mn}{2}   & 6122.434     & $ 0.950^a$ \nl
6122.81 &  1 &  \ion{Mn}{2}   & 6122.810     & $ 0.084^a$ \nl
6125.86 & 12 &  \ion{Mn}{2}   & 6125.863     & $ 0.783^a$ \nl
6126.22 &  6 &  \ion{Mn}{2}   & 6126.218     & $ 0.230^a$ \nl
6126.53 &  1 &  \ion{Mn}{2}   & 6126.51\phn  & $-0.791^a$ \nl
6128.74 & 10 &  \ion{Mn}{2}   & 6128.734     & $ 0.588^a$ \nl
6129.04 &  7 &  \ion{Mn}{2}   & 6129.033     & $ 0.208^a$ \nl
6129.26 &  1 &  \ion{Mn}{2}   & 6129.254     & $-0.792^a$ \nl
6130.80 & 10 &  \ion{Mn}{2}   & 6130.796     & $ 0.354^a$ \nl
6131.02 &  6 &  \ion{Mn}{2}   & 6131.016     & $ 0.053^a$ \nl
6131.93 &  6 &  \ion{Mn}{2}   & 6131.923     & $ 0.053^a$ \nl
6138.78 &  4 &  \ion{P}{2}    & 6138.74\phn  & $-1.31^b$ \nl
        &    &  \ion{P}{2}    & 6138.78\phn  & $ 0.12^b$ \nl
6139.53 &  1 &  \nodata       & 6139.53\phn  &\nodata \nl
6149.46 &  3 &  \ion{Hg}{2}   & 6149.475     & $ 0.328^a$ \nl
6152.52 &  3 &  \ion{P}{2}    & 6152.51\phn  & $-0.01^b$ \nl
6155.50 &  1 &  \ion{P}{2}    & 6155.50\phn  &\nodata \nl
6165.93 &  1 &  \ion{P}{2}??  & 6165.60\phn  & $-0.469^a$ \nl
6168.53 &  3 &  \ion{P}{2}    & 6168.53\phn  & $ 0.12^b$ \nl
6169.15 &  1 &  \ion{P}{2}    & 6169.12\phn  & $-1.44^b$ \nl
        &    &  \ion{P}{2}    & 6169.17\phn  & $-2.87^b$ \nl
6170.92 &  1 &  \nodata       & \nodata      &\nodata \nl
6172.64 &  2 &  \ion{P}{2}?   & 6172.80\phn  &\nodata \nl
6180.02 &  2 &  \nodata       & \nodata      &\nodata \nl
6180.56 &  4 &  \ion{P}{2}    & 6180.54\phn  & $ 0.22^b$ \nl
        &    &  \ion{P}{2}    & 6180.68\phn  & $-1.42^b$ \nl
6361.79 &  5 &  \ion{P}{2}    & 6361.80\phn  & $ 0.20^b$ \nl
6375.79 &  3 &  \nodata       & \nodata      &\nodata \nl
6401.09 &  4 &  \ion{P}{2}    & 6401.02\phn  & $-1.38^b$ \nl
        &    &  \ion{P}{2}    & 6401.08\phn  & $ 0.05^b$ \nl
6413.77 &  3 &  \ion{P}{2}    & 6413.75\phn  & $-0.08^b$ \nl
\enddata
\tablecomments{Wavelengths are for air.
Sources for $\log(gf)$ values are: (a) Kurucz \& Bell (1995), (b) Coulomb approximation.}
\end{deluxetable}

\end{document}